\documentclass{Interspeech}
\usepackage{multirow}
\usepackage{booktabs} 
\usepackage{threeparttable} 
\usepackage{amsmath}

\interspeechcameraready

\title{Cryfish: On deep audio analysis with Large Language Models}

\author[affiliation={1,2}]{Anton}{Mitrofanov}
\author[affiliation={1,2}]{Sergei}{Novoselov}
\author[affiliation={1}]{Tatiana}{Prisyach}
\author[affiliation={1,2}]{Vladislav}{Marchevskiy}
\author[affiliation={1}]{Arseniy}{Karelin} 
\author[affiliation={1,2}]{Nikita}{Khmelev}
\author[affiliation={1,2}]{Dmitry}{Dutov}
\author[affiliation={1}]{Stepan}{Malykh}
\author[affiliation={1}]{Igor}{Agafonov}
\author[affiliation={2}]{Aleksandr}{Nikitin}
\author[affiliation={1}]{Oleg}{Petrov}

\affiliation{}{Speech Technology Center}{Russia}
\affiliation{}{ITMO University}{Russia}
\email{mitrofanov-aa,novoselov,prisyach,marchevskiy,karelin,khmelev,dutov,malykh-s,agafonov-i @speechpro.com, nikitin@itmo.ru, petrov-o@speechpro.com}

\newcommand{\mycomment}[1]{}
\usepackage{comment}

\begin{document}

\maketitle

\begin{abstract}

    The recent revolutionary progress in text-based large language models (LLMs) has contributed to the growth of interest in extending capabilities of such models to multimodal perception and understanding tasks. Hearing is an essential capability that is highly desired to be integrated into LLMs. However, effective integrating listening capabilities into LLMs is a significant challenge lying in generalizing complex auditory tasks across speech and sounds. To address these issues, we introduce Cryfish, our version of auditory-capable LLM. The model integrates WavLM audio-encoder features into Qwen2 model using a transformer-based connector. Cryfish is adapted to various auditory tasks through a specialized training strategy. We evaluate the model on the new Dynamic SUPERB Phase-2 comprehensive multitask benchmark specifically designed for auditory-capable models. The paper presents an in-depth analysis and detailed comparison of Cryfish with the publicly available models.

\end{abstract}

\section{Introduction}


In recent years, the field of artificial intelligence has witnessed remarkable advancements with the development of LLMs. 
While the latest models demonstrate impressive capabilities in text processing, instruction following and task solving, there is a growing interest in extending their capabilities to multiple modalities \cite{team2023gemini, wu2023next}. 
Recent works, including  the development of SALMONN \cite{tang2024salmonn}, WavLLM \cite{hu2024wavllm}, and Qwen2-Audio \cite{chu2024qwen2}, have made significant progress in incorporating audio understanding into LLMs. 
For example, researchers have extended LLMs to process speech using pretrained encoders like Whisper \cite{radford2023robust}, aligning audio features with text embeddings \cite{chu2024qwen2}. 
To avoid explosion of context length from numerous audio frames, compression methods like Q-Former \cite{tang2024salmonn}
are employed.
Compared to speech, audio event processing requires specialized approaches. There are works on using speech encoders to capture paralinguistic features \cite{gong2023whisper}, 
but these tagging methods lack temporal resolution by design. 
Another approach is to connect audio event encoders directly to LLMs \cite{liu2024music}, similar to the already mentioned speech case to not lose temporal information.

However, existing auditory-capable LLMs (AudioLLMs) 
demonstrate limited generalization capabilities across diverse audio-related tasks
showing significant performance degradation on tasks different from their training distribution \cite{huang2024dynamic}.
Therefore, our goal was to improve AudioLLM generalization over a wide range of audio-related tasks. 
Our work addresses this limitation by developing a comprehensive approach to audio understanding in LLMs, focusing on enhancing their ability to generalize across diverse audio processing tasks.


To evaluate the effectiveness of our approach, we utilize the recently introduced Dynamic SUPERB Phase-2 benchmark \cite{huang2024dynamic}, which provides a comprehensive framework for assessing AudioLLMs across diverse tasks. We extend our evaluation beyond the benchmark by analyzing the model's performance on specific challenging tasks such as language identification and speaker verification. These experiments not only demonstrate the capabilities of our system but also provide insights into the generalization abilities of AudioLLMs.

In this paper, we present Contextual Response Yielding Framework for Intelligent Sound Handling (Cryfish).
We describe approaches for integrating classical LLMs with audio modality and evaluate AudioLLM's capability to solve tasks beyond its training data scope.
The contributions of our work are as follows:

\begin{itemize}
    \item A comprehensive data preparation pipeline that combines template-based and LLM-generated instructions to maintain both task-specific performance and NLP capabilities.
    \item An efficient architecture that integrates WavLM audio encoder with Qwen2 LLM through a transformer-based connector, enabling effective processing of various audio tasks.
    \item Evaluation of AudioLLMs on Dynamic SUPERB Phase-2 benchmark, demonstrating competitive performance against existing solutions.
    
    \item In-depth analysis of specific challenging tasks like speaker verification and language identification, providing insights into the model's capabilities and limitations.
\end{itemize}

\section{Training data preparation}
Training AudioLLM requires carefully prepared instruction-based data, where each training example consists of three key components: a natural language request (task description or question), one or more audio files (or none for text-only tasks), and the expected response in natural language.

Our model training procedure consists of two main stages: template-based instruction training and LLM-generated instruction training. This dual-stage approach allows us to first establish strong baseline capabilities using structured templates, then expand to more natural language interactions through generated instructions.
\subsection{Template instructions}
We used an extensive collection of audio-text template instructions, totalling 13k hours of audio recordings. 
For each speech corpus, we saved all the metadata and used it in the instructions. In general, we managed to collect data on the gender, age of the speakers, texts in different languages, audio descriptions, and language identifiers. For some datasets, additional information was provided partially. For those datasets, we removed all partially annotated samples.

We expanded our training data by augmenting the LibriSpeech \cite{Librispeech} database to include additional classification tasks for noise, reverberation, distance and signal-to-noise ratio (SNR). We used television noise \cite{richey2018voices}, noise from the MUSAN corpus \cite{MUSAN} and synthetic RIRs from \cite{SimulatedRIRs}. The noises were added to the reverberated or clean speech signal with different SNR and expanded by repetition to cover the entire audio file.

For the audio editing detection task, we also augmented LibriSpeech \cite{Librispeech}. We defined a set of edits (cuts, pitch shifting, volume edits), then for each audio we randomly chose 1-4 edits to perform along with regions where to do each edit (beginning, middle or end section of an audio). The questions were formulated in a chain-of-thought (COT) manner, where each edit was used as a "reason" for COT in templates. We also balanced the dataset with negative responses, using the lack of specific editing techniques as COT reasons.

The information about template instruction datasets and the number of instructional samples is presented in Table~\ref{table:main-instr-set}.

\begin{table}[!htb]
\begin{minipage}[t]{0.5\textwidth}
 \begin{threeparttable}[b]
\centering
\caption{\label{table:main-instr-set} Template instruction datasets.}
\begin{tabular}{|l|l|l|}
\hline
\textbf{Corpus} & \textbf{Task} & \textbf{Size}\\ \hline \hline
Auto-ACD \cite{Auto-ACD}  & audiocaption & 1680k \\ 
WavCaps \cite{WavCaps} & audiocaption, QA & 540k \\
MLS \cite {MLS} & speech \& lang & 850k \\
Fleurs \cite{FLEURS}  & speech \& lang & 270k  \\
Chime8-train \cite{CHiME-6} & speech &  404k\\
Common Voice\cite{commonvoice:2020}  & age, gender, speech, lang & 644k \\
Librispeech\tnote{1} \cite{Librispeech} & speech, gender, SNR, dist & 122k  \\
Voxceleb2\tnote{2} \cite{VoxCeleb2}  & speaker verification & 146k \\
Phoneme text db\tnote{3}  & ARPA g2p, IPA g2p & 314k  \\
\hline
\end{tabular} 
    \begin{tablenotes}
       \item [1] \textit{Augmentation with RIRs and noises.}
       \item [2] \textit{Each instruction pairs two audio segments with speaker identity verification and ID requests.}
       \item [3] \textit{Text-only data for ARPA and IPA phonetic transcription.}
     \end{tablenotes}
 \end{threeparttable}
\end{minipage} 
\vspace{-5mm}
\end{table}

\subsection{Instruction Generation}
%


AudioLLM systems face dual challenges: natural language understanding and audio signal processing. 
While template-based instruction training provides structured audio processing capabilities, it demonstrates limitations in handling unconstrained user requests.
To address these challenges systematically, we developed a specialized instruction generation pipeline.

We utilized dataset descriptions and metadata annotations to generate audio-specific instructions, with source datasets detailed in Table~\ref{table:add-instr-set}. 

For each audio sample, we combined dataset descriptions with metadata tags (e.g., instrument types, moods) into structured prompts\footnote{Specifics of the process can be found in https://github.com/medbar/cryfish}, and used text based LLMs to synthesize instructions based on all known about the audio info.
Using Qwen2.5 models \cite{yang2024qwen2}, we generated question-answer pairs which maintained stylistic consistency with the LLM inside AudioLLM.


%


\begin{table}[!htb]
\begin{minipage}[t]{0.5\textwidth}
 \begin{threeparttable}[b]
\centering
    \caption{\label{table:add-instr-set} Instruction Generation datasets.}
\begin{tabular}{|l|l|l|}
\hline
\textbf{Task} & \textbf{Corpus} & \textbf{Size}\\ \hline \hline
Speech & SpokenSTS 
, CoVoST-2 
       \cite{wang2020covost} & 160k \\
       & Emilia \cite{emilia}, Voxceleb2\tnote{1}, Common Voice, & \\
       & Speech-MASSIVE \cite{speechmassiv},  FSDD \cite{jackson2018jakobovski} & \\
       & Dailytalk \cite{lee2022dailytalk}, Ara-eng-cs \cite{rashad2024arabicwhisper}  & \\ \hline
Audio & Auto-ACD, Spatial LibriSpeech \cite{spatial_librispeech2023} & 40k \\ \hline
Music & Mridangam \cite{Mridangam}, OpenMIC \cite{OpenMIC}, & 104k \\
      & Vocalset \cite{Vocalset}, GTZAN \cite{GTZAN}, & \\
      & NSynth \cite{nsynth2017}, MTG-Jamendo \cite{bogdanov2019mtg} & \\ \hline
Safety &  MLAAD \cite{muller2024mlaad},  Multilingual-tts \cite{multilingual-tts} & 143k \\
       & CtrSVDD \cite{CtrSVDD}, SceneFake \cite{SceneFake} & \\
       & MUSDB18-HQ \cite{MUSDB18-HQ}, TTS\_refusal\tnote{2} & \\ \hline
Emotion & CREMA-D \cite{CREMA-D}, IEMOCAP \cite{IEMOCAP}, & 36k \\
        & MSP-Podcast \cite{MSP-Podcast}, & \\ \hline
Prosody & SpeechAccent \cite{Speech-accent-archive}, Speechocean \cite{zhang2021speechocean762} & 7k \\ \hline
Other & small databases on human health, & 0.7k \\
& data for anomalous sound detection, & \\
& bird databases & \\ \hline
\end{tabular}
    \begin{tablenotes}
       \item [1] \textit{Diarization and predicting speakers timestamps.}
       \item [2] \textit{Real voices with noTTS responses.}
     \end{tablenotes}
 \end{threeparttable}
\end{minipage}
\vspace{-5mm}
\end{table}

Despite generating linguistically diverse instructions, the approach showed limitations: bias toward binary questions, inconsistent metadata tag utilization, and generation of metadata-focused requests bypassing audio content. We addressed these issues through regex-based filtering and template fallbacks. 

To evaluate the quality of the instruction generation method, we conducted a manual assessment on 250 randomly sampled instructions. The observed accuracy was 0.795, indicating acceptable instructional accuracy. 





\section{Model architecture}

Cryfish architecture was inspired by SALMONN \cite{tang2024salmonn}, which consists of two types of audio encoders, connector layers, and LLM with Low-Rank Adaptation (LoRA) \cite{hu2021lora} adapter. Audio encoders are an essential part of the model as they provide a rich variety of audio features used for auditory tasks solving. 
However, supervised-learned (SL) encoders limit not only the model's ability to handle tasks that are not covered by their pre-training objectives but also constrain the context length (e.g., Whisper is limited to 30 seconds).


WavLM \cite{chen2022wavlm}, trained on a large corpus of self-supervised data, provides rich audio representations that can potentially handle a wide range of tasks, including those outside typical supervised training objectives. 
Using a single universal encoder simplifies the experimental setup and provides a clearer path for future improvements, as it eliminates the complexity of managing multiple specialized encoders.

To better capture sentence-level information when bridging these audio features with the language model, we replace the windowed Q-Former with a transformer-based connector. 
This connector extracts 5 sentence-level embeddings and a sequence of frame-level embeddings at a rate of 2.5 Hz. The connector remains trainable during the training process, while the WavLM extractor was frozen until the last epoch.

The embeddings generated by the connector are integrated with text instruction prompts and passed to the Qwen-2.5-7B-Instruct \cite{yang2024qwen2}. 
The connector and WavLM together comprise 345M parameters, and the language model adapter contains 21M parameters.


\section{Training process}

We adopted a two-stage training strategy. Stage 1 involved pre-training on a large-scale audio-text dataset to develop fundamental audio understanding capabilities. In Stage 2, we fine-tuned the model on diverse instruction data to enable task-specific responses.
\label{sec:training:stage1}
In Stage 1, we focused on establishing baseline audio processing capabilities through training the connector and adapting the language model. This initial stage utilized 3.2M template-based instructions across 7 tasks: ASR, Gender recognition, Speaker identification, Language identification, Noise classification, Distance and SNR predictions. Training ran for two epochs, one with frozen and one with unfrozen WavLM, using Adam optimizer with OneCycleLR scheduler (max learning rate 1e-3) and dynamic batch size around 3000 tokens per batch.
\label{sec:training:stage2}
Stage 2 involved training on instruction data generated through LLM, enhancing the model's ability to handle diverse and complex audio-related tasks. We utilized 44 different datasets, totaling 0.8M instructions (0.6M generated and 0.2M template-based), maintaining the same training settings as Stage 1.



\section{Results}
To evaluate AudioLLM systems, researchers can choose from various specialized benchmarks: AudioBench \cite{wang2024audiobench} for speech-language understanding, MuChoMusic \cite{weck2024muchomusic} for music processing, and frameworks covering multiple domains like AIR-Bench \cite{yang2024air} and Dynamic-SUPERB Phase-2 (DSB) \cite{huang2024dynamic}. DSB offers a unique advantage due to its diverse set of 100+ tasks, and despite having fewer examples per task ($\sim$ 200 samples), its broad coverage makes it particularly suitable for evaluating model generalization. 

For the response generation process, we used the best path search with beam width 3, repetition penalty 1.1 and max tokens 500.

\subsection{Dynamic-SUPERB Phase-2}
For evaluating classification tasks in DSB, we employed the LLM-as-a-judge methodology. After comparing several judge models (GPT-4, GPT-3.5, LLaMA 3.3, and DeepSeek), LLaMA 3.3 70B demonstrated superior performance with 97.78\% accuracy (2468 correct assessments out of 2524) on a balanced human-evaluated dataset. For regression tasks, we extracted task-specific information directly from the model's output, except for LibriTTS\_PoS and LibriTTS\_PoS\_with\_transcription tasks, which required LLM post-processing.

For the comparative analysis across different models, we evaluated 154 DSB tasks. Table~\ref{table:Domain-level-results} presents the most interpretable regression tasks (first five rows), including speech-to-text translation (SuperbST, English to German), multilingual ASR (MLS, Italian and Polish), and distance prediction (Audio Spatial Distance Prediction, measured by Median Absolute Error). The N/A rate (NAR) metric quantifies request comprehension failures. The Average Speech LLM-C and Average Audio LLM-C represent the mean LLM-as-a-judge accuracy in speech and audio domains, respectively.

\begin{figure}[t]
\centering
\includegraphics[width=\linewidth]{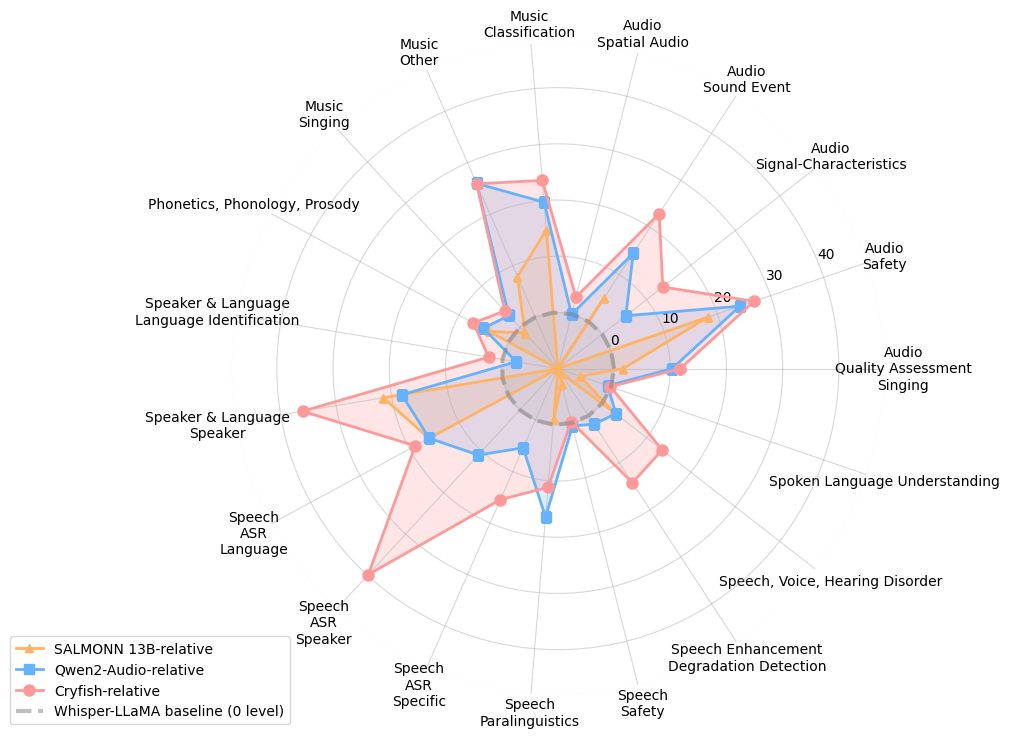}
\caption{Performance comparison on the Dynamic SUPERB benchmark Phase-2. 
The figure presents relative-score-based performances of AudioLLMs across 19 domains, normalized with respect to the Whisper-LLaMA baseline.}
\label{fig:dsb_results}
\vspace{-5mm}
\end{figure}

\begin{table*}[!t]
\caption{Comparative analysis of AudioLLM models on Dynamic SUPERB Phase-2 tasks, showing representative regression metrics (BLEU, WER, MEDAE), classification accuracy (LLM-C), and relative-score-based performances across speech and audio domains.}
  \label{table:Domain-level-results}
\vspace{-1mm}
\centering
\begin{tabular}{|c|c|c|c|c|c|c|}
\hline
\textbf{Task or Domain} & \textbf{Metric} & \textbf{Whisper-LLaMA} & \textbf{SALMONN 7B} & \textbf{SALMONN 13B} & \textbf{Qwen2-Audio} & \textbf{Cryfish} \\ \hline
\hline
SuperbST & BLEU ↑ & 17.37 & 18.79 & 16.78 & 23.35 & \textbf{35.36} \\ \hline
MLS it & WER ↓ & 85.38 & 39.11 & 31.54 & 33.16 & \textbf{29.55} \\ \hline
MLS pl & WER ↓ & 76.79 & 73.83 & 51.53 & 101.41 & \textbf{46.32} \\ \hline
SUPERB ASR & WER ↓ & 33.96 & 15.11 & \textbf{2.79} & 36.70 & 5.68 \\ \hline
\begin{tabular}[c]{@{}c@{}}Audio Spatial \\ Distance Prediction\end{tabular} & \begin{tabular}[c]{@{}c@{}}NA-adj medae ↓\\ medae↓/(1-NAR↓)\end{tabular} & \begin{tabular}[c]{@{}c@{}}1.34 \\ 0.67/0.50\end{tabular} & \begin{tabular}[c]{@{}c@{}} 1.18\\ \textbf{0.45}/0.38\end{tabular} & \begin{tabular}[c]{@{}c@{}}-\\ 0.00/0.00\end{tabular} & \begin{tabular}[c]{@{}c@{}} 1.64\\ 0.69/0.42\end{tabular} & \begin{tabular}[c]{@{}c@{}} \textbf{0.84} \\ 0.76/\textbf{0.90}\end{tabular} \\ \hline
\hline
Speech & Avg LLM-C ↑ & 41.34 & 34.26 & 35.42 & 47.23 & \textbf{51.57} \\ \hline
Audio & Avg LLM-C ↑ & 15.07 & 22.70 & 24.99 & 30.43 & \textbf{37.71} \\ \hline
\hline
Speech & \begin{tabular}[c]{@{}r@{}}rel-score-based ↑\end{tabular} & 0 & -7.07 & -6.37 & 5.88 & \textbf{10.22} \\ \hline
Audio & \begin{tabular}[c]{@{}r@{}}rel-score-based ↑\end{tabular} & 0 & 7.62 & 9.91 & 15.35 & \textbf{22.64} \\ \hline
\hline
Total & \begin{tabular}[c]{@{}r@{}}rel-score-based ↑\end{tabular} & 0 & -2.78 & -1.57 & 8.65 & \textbf{13.85} \\ \hline
\end{tabular}
\vspace{-5mm}
\end{table*}
Due to heterogeneous evaluation metrics across tasks, a relative-score-based methodology was utilized for comprehensive assessment. This approach incorporates NAR by scaling base task metrics for regression tasks (multiplication or division, depending on metric direction). Table~\ref{table:Domain-level-results} presents relative-score-based performances for speech and audio domains, along with the total average relative-score-based performance. Figure~\ref{fig:dsb_results} shows AudioLLM performance comparison across 19 DSB domains, where Cryfish demonstrates superior performance in both speech and audio domains.
To ensure reproducibility and transparency, detailed evaluation results for each dataset, along with the evaluation and aggregation procedure, can be found in the project repository\footnote{https://github.com/medbar/cryfish}.

To complement the comprehensive cross-domain evaluation of Dynamic SUPERB Phase-2 benchmark, we conducted additional task-specific analyses to compare our model against specialized solutions.

\subsection{Speaker Verification}

To deepen our understanding of Cryfish speaker recognition (SR) performance, in addition to VC1-DSB (SuperbSV) we introduced other benchmarks: VC1O (a subsample from VoxCeleb1-O \cite{Nagrani19}), VOiCES (a subsample from VOiCES evaluation set \cite{nandwana19_interspeech}). The samples contain 2000 comparisons with equal representation of target and impostor pairs. We opted for well-established Equal Error Rate (EER) as our primary SR metric, which we calculated using logprob of "Yes"/"No" tokens.
Table~\ref{tab:sr_results} presents the results of the models:  Cryfish (our main model), Cryfish-stage1 (see \ref{sec:training:stage1}), Cryfish-vc2tuned (the model trained only on VC2 data after stage 1 training). For comparison, we included metrics of SALMONN-7B \cite{tang2024salmonn}, one of the leading AudioLLMs in speaker verification, and the prominent SR model ECAPA-TDNN \cite{desplanques20_interspeech}.

All variants of the Cryfish model outperform SALMONN-7B, with the most significant difference observed in the out-of-domain VOiCES evaluation protocol, where SALMONN shows near-random performance under noisy far-field conditions.
The best Cryfish variant in terms of SR is Cryfish-vc2tuned, 
 demonstrating the benefits of single-task training.
When evaluated with DSB benchmark prompts, all AudioLLMs showed sensitivity to task formulation, resulting in performance changes. The exception was Cryfish-vc2tuned, which, being overfitted to verification tasks, bypassed prompt understanding completely.

While Cryfish demonstrates superior performance compared to the SALMONN's results, its SR capabilities remain significantly below the task-specific ECAPA-TDNN baseline model.
\begin{table}[h!]
\caption{EER (\%) ↓ on different datasets. Values are obtained with prompts similar to the ones, used in a training stage. Values in parentheses are attained with benchmark-specific prompts}
\label{tab:sr_results}
\centering
\begin{tabular}{lccc}
\hline
\multicolumn{1}{|l|}{\textbf{Model}} & \multicolumn{1}{c|}{\textbf{VC1-DSB}} & \multicolumn{1}{c|}{\textbf{VC1O}} & \multicolumn{1}{c|}{\textbf{VOiCES}} \\ \hline \hline
\multicolumn{1}{|l|}{SALMONN-7B} & \multicolumn{1}{c|}{8.5 (12)} & \multicolumn{1}{c|}{10} & \multicolumn{1}{c|}{50} \\ \hline
\multicolumn{1}{|l|}{Cryfish} & \multicolumn{1}{c|}{\textbf{7.5} (\textbf{9})} & \multicolumn{1}{c|}{\textbf{9.9}} & \multicolumn{1}{c|}{\textbf{29.2}} \\ \hline
\hline
\multicolumn{1}{|l|}{Cryfish-stage1} & \multicolumn{1}{c|}{6.5 (13)} & \multicolumn{1}{c|}{10} & \multicolumn{1}{c|}{28.4} \\ \hline
\multicolumn{1}{|l|}{Cryfish-vc2tuned} & \multicolumn{1}{c|}{5 (3)} & \multicolumn{1}{c|}{6.7} & \multicolumn{1}{c|}{25.7} \\ \hline
\hline
\multicolumn{1}{|l|}{ECAPA-TDNN} & \multicolumn{1}{c|}{\textbf{2}} & \multicolumn{1}{c|}{\textbf{1.12}} & \multicolumn{1}{c|}{\textbf{4.83}} \\ \hline
\end{tabular}
\vspace{-5mm}
\end{table}

\subsection{Language Identification}
For the language identification evaluation, we used FLEURS \cite{FLEURS} (102 languages) and VoxLingua107 \cite{VoxLingua107} (33 languages) datasets. 
We compared Cryfish's language detection capabilities against classic multilingual models and existing AudioLLMs.
The evaluation included two protocols for AudioLLMs: open-ended language identification and closed-set classification with 6 options (one correct and five randomly selected languages). The results are presented in Table~\ref{tab:li_results}.
Cryfish outperforms the other considered AudioLLMs in language detection, particularly with the closed instruction set. 
While Qwen2-audio achieves high accuracy (98.7\%) on its supported languages (Chinese, English, Japanese, Korean, German, Spanish, and Italian), it shows significantly lower performance on other languages from FLEURS and VoxLingua107 even in the closed-set scenario. In contrast, Cryfish maintains robust performance across a comprehensive language set, approaching XLS-R and surpassing Whisper in the closed-set evaluation protocols.

\begin{table}[h]
\caption{Language identification accuracy (\%) ↑ on the test sets covering 33 and 102 languages.}
\label{tab:li_results}
\label{tab:lang_id}
\centering
\begin{tabular}{|l|ll|ll|}
\hline
\multirow{2}{*}{\textbf{Model}} & \multicolumn{2}{c|}{\textbf{VoxLingua107}} & \multicolumn{2}{c|}{\textbf{FLEURS}} \\
& Open & Closed & Open & Closed \\ \hline \hline
SALMONN-13B     & \multicolumn{1}{l|}{23.0} & 23.4 & \multicolumn{1}{l|}{2.5}                        & 5.6 \\ \hline
Qwen-audio  & \multicolumn{1}{l|}{29.8} & 36.4 & \multicolumn{1}{l|}{8.0} 
                & 22.2 \\ \hline
Qwen2-audio & \multicolumn{1}{l|}{47.2} & 56.6 & \multicolumn{1}{l|}{15.2}                       & 35.3 \\ \hline
Cryfish     & \multicolumn{1}{c|}{\textbf{66.2}} & \textbf{83.4} & \multicolumn{1}{l|}{\textbf{67.4}}                       & \textbf{89.3} \\ \hline
\hline
Whisper \cite{radford2023robust}    & \multicolumn{2}{c|}{-}                     & \multicolumn{2}{c|}{64.5}                                        \\ \hline
MMS \cite{MMS}         & \multicolumn{2}{c|}{\textbf{95.6}}                  & \multicolumn{2}{c|}{\textbf{89.6}}                                        \\ \hline
XLS-R \cite{babu2021xls}       & \multicolumn{2}{c|}{94.3}                  & \multicolumn{2}{c|}{-}                                        \\ \hline
\end{tabular}
\vspace{-5mm}
\end{table}



\subsection{Discussion}

Training revealed two critical challenges. First, the prompt-class distribution imbalance led to model decisions based on the prompt structure rather than the audio content, necessitating balanced (class, prompt-style) pairs. 
Second, shorter answer tasks produced lower cross-entropy loss compared to sequence generation tasks, requiring loss adjustment through masking or alternative loss functions for sequence generation tasks.

Additionally, the WavLM-based model showed better training stability than the Whisper+Beats+ECAPA combination. Despite the latter's potential advantages in feature extraction, we were unable to achieve comparable performance with this multi-encoder approach.

\section{Conclusions}

In this paper, we presented Cryfish, a novel AudioLLM that demonstrates strong performance across diverse audio processing tasks.
We proposed an efficient architecture that integrates WavLM with Qwen2 LLM through a transformer-based connector, and a comprehensive data preparation pipeline that combines template-based and LLM-generated instructions to maintain both task-specific performance and NLP capabilities.

Our experiments on the comprehensive multitask Dynamic-SUPERB Phase-2 and on a number of task-specific benchmarks showed that Cryfish outperforms existing open-source AudioLLMs on several key metrics. 
The results demonstrated that our approach effectively balances specialized task performance and general audio understanding capabilities. However, challenges remain in achieving parity with task-specific models while maintaining the flexibility of a general-purpose system. 


\section{Acknowledgements}
This work was supported by ITMO University (grant "Research work in the field of artificial intelligence", project No. 640110 "Voice personification for artificial intelligence systems").

\bibliographystyle{IEEEtran}
\bibliography{mybib}

\begin{thebibliography}{10}
\providecommand{\url}[1]{#1}
\csname url@samestyle\endcsname
\providecommand{\newblock}{\relax}
\providecommand{\bibinfo}[2]{#2}
\providecommand{\BIBentrySTDinterwordspacing}{\spaceskip=0pt\relax}
\providecommand{\BIBentryALTinterwordstretchfactor}{4}
\providecommand{\BIBentryALTinterwordspacing}{\spaceskip=\fontdimen2\font plus
\BIBentryALTinterwordstretchfactor\fontdimen3\font minus
  \fontdimen4\font\relax}
\providecommand{\BIBforeignlanguage}[2]{{%
\expandafter\ifx\csname l@#1\endcsname\relax
\typeout{** WARNING: IEEEtran.bst: No hyphenation pattern has been}%
\typeout{** loaded for the language `#1'. Using the pattern for}%
\typeout{** the default language instead.}%
\else
\language=\csname l@#1\endcsname
\fi
#2}}
\providecommand{\BIBdecl}{\relax}
\BIBdecl

\bibitem{team2023gemini}
G.~Team, R.~Anil \emph{et~al.}, ``Gemini: a family of highly capable multimodal
  models,'' \emph{arXiv preprint arXiv:2312.11805}, 2023.

\bibitem{wu2023next}
S.~Wu, H.~Fei \emph{et~al.}, ``Next-gpt: Any-to-any multimodal llm,''
  \emph{arXiv preprint arXiv:2309.05519}, 2023.

\bibitem{tang2024salmonn}
\BIBentryALTinterwordspacing
C.~Tang, W.~Yu \emph{et~al.}, ``Salmonn: Towards generic hearing abilities for
  large language models,'' 2024. [Online]. Available:
  \url{https://arxiv.org/abs/2310.13289}
\BIBentrySTDinterwordspacing

\bibitem{hu2024wavllm}
S.~Hu \emph{et~al.}, ``Wavllm: Towards robust and adaptive speech large
  language model,'' \emph{arXiv preprint arXiv:2404.00656}, 2024.

\bibitem{chu2024qwen2}
Y.~Chu, J.~Xu \emph{et~al.}, ``Qwen2-audio technical report,'' \emph{arXiv
  preprint arXiv:2407.10759}, 2024.

\bibitem{radford2023robust}
A.~Radford, J.~W. Kim \emph{et~al.}, ``Robust speech recognition via
  large-scale weak supervision,'' in \emph{PMLR}, 2023, pp. 28\,492--28\,518.

\bibitem{gong2023whisper}
Y.~Gong \emph{et~al.}, ``Whisper-at: Noise-robust automatic speech recognizers
  are also strong general audio event taggers,'' \emph{arXiv preprint
  arXiv:2307.03183}, 2023.

\bibitem{liu2024music}
S.~Liu \emph{et~al.}, ``Music understanding llama: Advancing text-to-music
  generation with question answering and captioning,'' in \emph{ICASSP
  2024-2024 IEEE International Conference on Acoustics, Speech and Signal
  Processing (ICASSP)}.\hskip 1em plus 0.5em minus 0.4em\relax IEEE, 2024, pp.
  286--290.

\bibitem{huang2024dynamic}
C.-y. Huang, W.-C. Chen \emph{et~al.}, ``Dynamic-superb phase-2: A
  collaboratively expanding benchmark for measuring the capabilities of spoken
  language models with 180 tasks,'' \emph{arXiv preprint arXiv:2411.05361},
  2024.

\bibitem{Librispeech}
V.~Panayotov, G.~Chen \emph{et~al.}, ``Librispeech: An asr corpus based on
  public domain audio books,'' in \emph{ICASSP}, 2015, pp. 5206--5210.

\bibitem{richey2018voices}
C.~Richey, M.~A. Barrios \emph{et~al.}, ``Voices obscured in complex
  environmental settings (voices) corpus,'' 2018.

\bibitem{MUSAN}
D.~Snyder \emph{et~al.}, ``Musan: A music, speech, and noise corpus,''
  \emph{arXiv:1510.08484v1}, 2015.

\bibitem{SimulatedRIRs}
T.~Ko \emph{et~al.}, ``A study on data augmentation of reverberant speech for
  robust speech recognition,'' in \emph{ICASSP}, 2017, p. 5220–5224.

\bibitem{Auto-ACD}
L.~Sun, X.~Xu \emph{et~al.}, ``Auto-acd: A large-scale dataset for
  audio-language representation learning,'' \emph{arXiv:2309.11500v4}, 2024.

\bibitem{WavCaps}
X.~Mei, C.~Meng, and et~al., ``Wavcaps: A chatgpt-assisted weakly-labelled
  audio captioning dataset for audio-language multimodal research,''
  \emph{arXiv:2303.17395v2}, 2024.

\bibitem{MLS}
V.~Pratap, Q.~Xu \emph{et~al.}, ``Mls: A large-scale multilingual dataset for
  speech research,'' \emph{ArXiv}, vol. abs/2012.03411, 2020.

\bibitem{FLEURS}
A.~Conneau and et~al., ``Fleurs: Few-shot learning evaluation of universal
  representations of speech,'' \emph{arXiv:2205.12446v1}, 2022.

\bibitem{CHiME-6}
S.~Watanabe and et~al., ``Chime-6 challenge: Tackling multispeaker speech
  recognition for unsegmented recordings,'' \emph{arXiv:2004.09249v2}, 2020.

\bibitem{commonvoice:2020}
R.~Ardila, M.~Branson \emph{et~al.}, ``Common voice: A massively-multilingual
  speech corpus,'' in \emph{LREC}, 2020, pp. 4211--4215.

\bibitem{VoxCeleb2}
J.~S. Chung \emph{et~al.}, ``Voxceleb2: Deep speaker recognition,''
  \emph{arXiv:1806.05622v2}, 2018.

\bibitem{yang2024qwen2}
A.~Yang, B.~Yang \emph{et~al.}, ``Qwen2. 5 technical report,'' \emph{arXiv
  preprint arXiv:2412.15115}, 2024.

\bibitem{wang2020covost}
C.~Wang \emph{et~al.}, ``Covost 2: A massively multilingual speech-to-text
  translation corpus,'' 2020.

\bibitem{emilia}
H.~He, Z.~Shang, and et~al., ``Emilia: An extensive, multilingual, and diverse
  speech dataset for large-scale speech generation,'' in \emph{Proc.~of SLT},
  2024.

\bibitem{speechmassiv}
B.~Lee \emph{et~al.}, ``Speech-massive: A multilingual speech dataset for slu
  and beyond,'' in \emph{Proc. Interspeech 2024}, 2024.

\bibitem{jackson2018jakobovski}
Z.~Jackson, C.~Souza \emph{et~al.}, ``Jakobovski/free-spoken-digit-dataset: v1.
  0.8,'' 2018.

\bibitem{lee2022dailytalk}
K.~Lee \emph{et~al.}, ``Dailytalk: Spoken dialogue dataset for conversational
  text-to-speech,'' 2022.

\bibitem{rashad2024arabicwhisper}
\BIBentryALTinterwordspacing
M.~Rashad, ``Arabic-whisper-codeswitching-edition,'' 2024. [Online]. Available:
  \url{https://huggingface.co/spaces/MohamedRashad/Arabic-Whisper-CodeSwitching-Edition}
\BIBentrySTDinterwordspacing

\bibitem{spatial_librispeech2023}
M.~Sarabia, E.~Menyaylenko \emph{et~al.}, ``Spatial librispeech: An augmented
  dataset for spatial audio learning,'' in \emph{Proc. Interspeech}, 2023, pp.
  3724--3728.

\bibitem{Mridangam}
A.~Anantapadmanabhan \emph{et~al.}, ``Modal analysis and transcription of
  strokes of the mridangam using non-negative matrix factorization,'' in
  \emph{ICASSP}, 2013, pp. 181--185.

\bibitem{OpenMIC}
E.~J. Humphrey \emph{et~al.}, ``Modal analysis and transcription of strokes of
  the mridangam using non-negative matrix factorization,'' in \emph{ISMIR},
  2018.

\bibitem{Vocalset}
J.~Wilkins, P.~Seetharaman \emph{et~al.}, ``Vocalset: A singing voice
  dataset,'' in \emph{ISMIR}, 2018.

\bibitem{GTZAN}
G.~Tzanetakis and P.~Cook, ``Musical genre classification of audio signals,''
  in \emph{IEEE Transactions on Speech and Audio Processing, vol. 10, no. 5},
  2002.

\bibitem{nsynth2017}
J.~Engel, C.~Resnick \emph{et~al.}, ``Neural audio synthesis of musical notes
  with wavenet autoencoders,'' 2017.

\bibitem{bogdanov2019mtg}
\BIBentryALTinterwordspacing
D.~Bogdanov, M.~Won \emph{et~al.}, ``The mtg-jamendo dataset for automatic
  music tagging,'' in \emph{ICML}, 2019. [Online]. Available:
  \url{http://hdl.handle.net/10230/42015}
\BIBentrySTDinterwordspacing

\bibitem{muller2024mlaad}
N.~M. M{\"u}ller, P.~Kawa \emph{et~al.}, ``Mlaad: The multi-language audio
  anti-spoofing dataset,'' \emph{IJCNN}, 2024.

\bibitem{multilingual-tts}
\BIBentryALTinterwordspacing
M.~Rashad, ``Multilingual-tts,'' 2023. [Online]. Available:
  \url{https://huggingface.co/datasets/MohamedRashad/multilingual-tts}
\BIBentrySTDinterwordspacing

\bibitem{CtrSVDD}
Y.~Zhang, Y.~Zang \emph{et~al.}, ``Svdd challenge 2024: A singing voice
  deepfake detection challenge (ctrsvdd track, training/development set),''
  \emph{https://zenodo.org/records/10467648}, 2024.

\bibitem{SceneFake}
J.~Yi, C.~Wang \emph{et~al.}, ``Scenefake: An initial dataset and benchmarks
  for scene fake audio detection,'' \emph{arXiv:2211.06073v2}, 2022.

\bibitem{MUSDB18-HQ}
\BIBentryALTinterwordspacing
Z.~Rafii, A.~Liutkus \emph{et~al.}, ``Musdb18-hq - an uncompressed version of
  musdb18,'' 2019. [Online]. Available:
  \url{https://doi.org/10.5281/zenodo.3338373}
\BIBentrySTDinterwordspacing

\bibitem{CREMA-D}
H.~Cao, D.~G. Cooper \emph{et~al.}, ``Crema-d: Crowd-sourced emotional
  multimodal actors dataset,'' in \emph{IEEE transactions on affective
  computing, 5(4)}, 2014, pp. 377--390.

\bibitem{IEMOCAP}
C.~Busso, M.~Bulut \emph{et~al.}, ``Iemocap: Interactive emotional dyadic
  motion capture database,'' in \emph{Journal of Language Resources and
  Evaluation, vol. 42, no. 4}, 2008, pp. 335--359.

\bibitem{MSP-Podcast}
R.~Lotfian and C.~Busso, ``Building naturalistic emotionally balanced speech
  corpus by retrieving emotional speech from existing podcast recordings,'' in
  \emph{IEEE Transactions on Affective Computing, vol. 10, no. 4}, 2019, pp.
  471--483.

\bibitem{Speech-accent-archive}
\BIBentryALTinterwordspacing
S.~Weinberger, ``Speech accent archive,'' 2015. [Online]. Available:
  \url{https://www.kaggle.com/datasets/rtatman/speech-accent-archive}
\BIBentrySTDinterwordspacing

\bibitem{zhang2021speechocean762}
J.~Zhang, Z.~Zhang \emph{et~al.}, ``speechocean762: An open-source non-native
  english speech corpus for pronunciation assessment,'' in \emph{Proc.
  Interspeech}, 2021.

\bibitem{hu2021lora}
E.~J. Hu, Y.~Shen \emph{et~al.}, ``Lora: Low-rank adaptation of large language
  models,'' \emph{arXiv preprint arXiv:2106.09685}, 2021.

\bibitem{chen2022wavlm}
S.~Chen \emph{et~al.}, ``Wavlm: Large-scale self-supervised pre-training for
  full stack speech processing,'' \emph{IEEE Journal of Selected Topics in
  Signal Processing}, vol.~16, no.~6, pp. 1505--1518, 2022.

\bibitem{wang2024audiobench}
B.~Wang \emph{et~al.}, ``Audiobench: A universal benchmark for audio large
  language models,'' \emph{arXiv preprint arXiv:2406.16020}, 2024.

\bibitem{weck2024muchomusic}
B.~Weck, I.~Manco, E.~Benetos \emph{et~al.}, ``Muchomusic: Evaluating music
  understanding in multimodal audio-language models,'' \emph{arXiv preprint
  arXiv:2408.01337}, 2024.

\bibitem{yang2024air}
Q.~Yang, J.~Xu \emph{et~al.}, ``Air-bench: Benchmarking large audio-language
  models via generative comprehension,'' \emph{arXiv preprint
  arXiv:2402.07729}, 2024.

\bibitem{Nagrani19}
A.~Nagrani, J.~S. Chung, W.~Xie, and A.~Zisserman, ``Voxceleb: Large-scale
  speaker verification in the wild,'' \emph{Computer Science and Language},
  2019.

\bibitem{nandwana19_interspeech}
M.~K. Nandwana, J.~van Hout, C.~Richey, M.~McLaren, M.~A. Barrios, and
  A.~Lawson, ``The voices from a distance challenge 2019,'' in
  \emph{Interspeech 2019}, 2019, pp. 2438--2442.

\bibitem{desplanques20_interspeech}
B.~Desplanques \emph{et~al.}, ``Ecapa-tdnn: Emphasized channel attention,
  propagation and aggregation in tdnn based speaker verification,'' in
  \emph{Interspeech}, 2020, pp. 3830--3834.

\bibitem{VoxLingua107}
T.~A. Jörgen~Valk, ``Voxlingua107: a dataset for spoken language
  recognition,'' \emph{arXiv:2011.12998v1}, 2020.

\bibitem{MMS}
V.~Pratap \emph{et~al.}, ``Scaling speech technology to 1,000+ languages,''
  \emph{arXiv:2305.13516v1}, 2023.

\bibitem{babu2021xls}
A.~Babu \emph{et~al.}, ``Xls-r: Self-supervised cross-lingual speech
  representation learning at scale,'' \emph{arXiv preprint arXiv:2111.09296},
  2021.

\end{thebibliography}

\end{document}